\documentclass[11pt]{amsart}
\usepackage{geometry}                
\usepackage{graphicx}
\usepackage{amssymb}
\usepackage{epstopdf}
\title{Quantum Wavefunction in a Schwarzschild Spacetime}
\author{Ikjyot Singh Kohli}
\date{October 27, 2011}                                           

\begin{document}
\maketitle

\begin{abstract}
A pseudo-Riemannian manifold contains an inherent Hamiltonian structure within the symplectic manifold in the cotangent bundle corresponding to the metric. Using this structure, it is possible to define a Hamiltonian, which can be quantized, and substituted into The Schrodinger Equation. The wave function, despite having a very complicated structure, has the very interesting property where it vanishes at the Schwarzschild $(r=0)$ singularity. This demonstrates an apparent violation of Quantum Mechanics, in which information is required to be conserved, and gives further evidence of a wave function vanishing at the singularity point. In addition, we give conditions for this wave function vanishing in terms of quantum numbers.
\end{abstract}

\newpage

\section{Introduction}

Our goal is to describe the classical Schrodinger wave function in the vicinity of a Schwarzschild singularity, specifically, at the $r=0$ physical singularity. It is a well known fact that from the Schwarzschild metric:

\begin{equation}
ds^2 = \left(1-\frac{2GM}{r}\right)dt^2 - \left(1-\frac{2GM}{r}\right)^{-1}dr^2 -r^2\left(d\theta^2 + \sin^2\theta d\phi^2\right)
\end{equation}
, the value of the Kretschmann scalar indicates a physical singularity at $r=0$:

\begin{equation}
K = R^{abcd}R_{abcd} = \frac{48G^2M^2}{r^6}
\end{equation}

The challenge is how one should introduce any type of quantum formalism in the context of general relativity. This can be done by realizing that the structure of the pseudo-Riemannian manifold, where such a manifold induces a linear relationship between the tangent and cotangent spaces, from which one can obtain a cometric tensor $g^{ab}$, where $g^{ab}$ is the inverse of the standard metric tensor $g_{ab}$.  There is a very clear relationship between the cometric tensor and the Hamiltonian on the symplectic manifold as explained in the next section. \footnote{Throughout, we use the convention that $c=1$.}

\section{Defining a Hamiltonian}

Following Jacobi's formulation of The Principle of Least Action \cite{Goldstein}, one can define a Hamiltonian in terms of the cometric tensor as follows:

\begin{equation}
H(q,p) = \frac{1}{2}g^{ij}p_{i}p_{j}
\end{equation}
, where $p_{i}$ denotes the standard canonical momentum in classical mechanics. This Hamiltonian, although, upon first glance seems to represent a free ``particle'', one can easily extend this definition of the Hamiltonian for a potential energy term in terms of the position coordinates ${q}$ as follows. If the Hamiltonian is of the general form:

\begin{equation}
H = \frac{1}{2}{M}^{ij}p_{i}p_{j} + V(q)
\end{equation}
, and one considers regions of the configuration space where $E-V(x) \neq 0$ (where $E$ denotes the total classical energy of the system, and is assumed to take on some fixed quantity), we can still derive a cometric of the form:

\begin{equation}
g^{ij} = 2(E-V)M^{ij}
\end{equation}

The above methodology, although, not specifically used in this paper, is an extension of The Maupertuis principle, and is extensively discussed in \cite{Arnold}.

Going back to Eq. (3), one can obtain a Hamiltonian by knowing the form of the metric tensor $g_{ij}$, which in this case is a solution of the Einstein field equations. We are specifically interested in The Schwarzschild solution. Instead of using the metric defined in Eq. (1) with the coordinate chart $(t,r,\theta,\phi)$, we will use isotropic rectangular coordinates \cite{Buchdahl}. \footnote{The reason for this is that The Schrodinger equation is slightly easier to solve with the Hamiltonian obtained from the latter rather than the former.} The specific coordinate transformations are:

\begin{equation}
x = r \sin\theta \cos\phi, y = r \sin\theta \sin\phi, z = r \cos\theta
\end{equation}

Here, $r$ takes the standard definition as: $r = \left(x^2+y^2+z^2\right)^{1/2}$. The isotropic rectangular Schwarzschild metric is now written as:
\begin{equation}
ds^2 = \frac{(1-GM/2r)^2}{(1+GM/2r)^2}dt^2 - (1+GM/2r)^4 (dx^2 + dy^2 + dz^2)
\end{equation}

For additional simplicity and the fact that The Schwarzschild solution is a static solution, we will use $3+1$ formulation to obtain an expression for the spatial 3-metric, $\gamma_{ij}$, and work with that:

\begin{equation}
\gamma_{ij} = -(1+GM/2r)^4(dx^2 + dy^2 + dz^2)
\end{equation}

It is interesting to note that the 3-metric $\gamma_{ij}$ is completely independent of time, so in terms of the ADM formalism, the evolution equation of the metric and extrinsic curvature tensors both vanish. Therefore, since the metric tensor is constant throughout any time evolution, it is Hamiltonian structure as defined above will remain constant as well. 

Applying Eq. (3) to the inverse of Eq. (8), we can see that the Hamiltonian takes the form:

\begin{equation}
H = -\frac{1}{2}(1+GM/2r)^{1/4}(p_x^2 + p_y^2 + p_z^2)
\end{equation}

This the Hamiltonian we will use in solving Schrodinger's equation in what follows.


\section{Quantizing The Hamiltonian and Solving Schrodinger's Equation}
Quantizing the Hamiltonian in Eq. (9), its position-space representation is:

\begin{equation}
\hat{H} = \frac{1}{2}(1+GM/2r)^{1/4}(\hbar^2 \hat{\nabla}^2)
\end{equation}

 The time-independent Schrodinger equation corresponding to this Hamiltonian is:
\begin{equation}
\hat{\nabla}^2 \psi(\textbf{x}) = \frac{2E}{\hbar^2} \left(1+\frac{GM}{2r}\right)^4 \psi(\textbf{x})
\end{equation}

Because this equation doesn't separate in rectangular coordinates, we employ spherical coordinates for our representation of $\textbf{x}$, and obtain:
\begin{equation}
\hat{\nabla}^2 \psi(r,\theta,\phi)  = \psi(r,\theta,\phi)\frac{2E}{\hbar^2}\left(1+\frac{GM}{2r}\right)^4
\end{equation}

Substituting the definition of the Laplacian operator in spherical coordinates, Eq. (12) becomes:

\begin{equation}
\frac{1}{r^2}\frac{d}{dr}\left(r^2 \frac{d\psi}{dr}\right) + \frac{1}{r^2\sin\theta}\frac{d}{d\theta}\left(\sin\theta \frac{d\psi}{d\theta}\right) + \frac{1}{r^2\sin^2\theta}\frac{d^2\psi}{d\phi^2} = \psi(r,\theta,\phi) \frac{2E}{\hbar^2}\left(1+ \frac{GM}{2r}\right)^4
\end{equation}

Applying the standard separation of variables technique, where we seek solutions of the form:
\begin{equation}
\psi(r,\theta,\phi) = R(r)\gamma(\theta, \phi)
\end{equation}

We obtain first the equation for $\gamma(\theta,\phi)$, (the angular equation) as:
\begin{equation}
\frac{1}{\gamma} \left(\frac{1}{\sin\theta}\frac{d}{d\theta}\left(\sin\theta \frac{d\gamma}{d\theta}\right) + \frac{1}{\sin^2\theta}\frac{d^2\gamma}{d\phi^2}\right) = -l(l+1)
\end{equation}

This solutions to this differential equation are the well-known spherical harmonics:

\begin{equation}
\gamma^{m}_{l} = \sqrt{(2l+1)\frac{(l-m)!}{(l+m)!}} P^{m}_{l}(\cos\theta)\exp(im\phi)
\end{equation}

(Where, as usual, the $P^{m}_{l}$ are the associated Legendre polynomials.)

The equation for $R(r)$, the radial equation is found to be:

\begin{equation}
2r\frac{dR}{dr}+\frac{d^2R}{dr^2}r^2 - \left[l(l+1)+r^2\frac{2E}{\hbar^2}\left(1+\frac{GM}{2r}\right)^4\right]R(r)=0
\end{equation}

The $\left(1+\frac{GM}{2r}\right)^4$ makes this differential equation very difficult to solve. However, one can see that if we expand this expression:

\begin{equation}
\left(1+\frac{GM}{2r}\right)^4 = 1 + \frac{G^4M^4}{16r^4} + \frac{G^3M^3}{2r^3} + \frac{3G^2M^2}{2r^2} + \frac{2GM}{r}
\end{equation}

Exact solutions for $R(r)$ are possible neglecting terms on the order of $1/r^3$. 

Therefore we state that:

\begin{equation}
\left(1+\frac{GM}{2r}\right)^4 \approx 1 + \frac{2GM}{r} + \frac{3G^2M^2}{2r^2}
\end{equation}

The solution for this modified differential equation is found to be:

\begin{equation}
R(r) = \exp \left(\frac{-\sqrt{2E}r}{\hbar}\right) r^{\frac{1}{2}\left(\sqrt{1+4l+4l^2+\frac{12EG^2M^2}{\hbar^2}}-1\right)} \left[C_{1} U(i,j,k) + C_{2}L(d,e,f)\right]
\end{equation}

Note that $C_{1}, C_{2}$ are arbitrary constants, and $U(i,j,k), L(d,e,f)$ are the confluent hypergeometric and generalized Laguerre polynomial functions respectively. The confluent hypergeometric function is defined as:

\begin{equation}
U(i,j,k) = \frac{1}{\Gamma(i)} \int^{\infty}_{0} \exp(-k t) t^{i-1} (1+t)^{j-i-1} dt
\end{equation}

Also, the parameters denoted above are defined to be:
\begin{equation}
i = \frac{1}{2}\left(1+ \frac{2\sqrt{2E}GM}{\hbar} + \sqrt{1+4l+4l^2+ \frac{12EG^2M^2}{\hbar^2}}\right)
\end{equation}

\begin{equation}
j = 1 + \sqrt{1+4l+4l^2 + \frac{12EG^2M^2}{\hbar^2}}
\end{equation}

\begin{equation}
k = \frac{2\sqrt{2E}r}{\hbar}
\end{equation}

\begin{equation}
d = -\frac{\hbar + 2\sqrt{2E}GM + \hbar \sqrt{1+4l+4l^2 + \frac{12EG^2M^2}{\hbar^2}}}{2\hbar}
\end{equation}

\begin{equation}
e = \sqrt{1+4l+4l^2+ \frac{12EG^2M^2}{\hbar^2}}
\end{equation}

\begin{equation}
f = \frac{2\sqrt{2E}r}{\hbar}
\end{equation}

We now wish to evaluate Eq. (20) at $r=0$, that is, at the location of the singularity point. We see that $R(0) = 0$ if:

\begin{equation}
r^{\frac{1}{2}\left(\sqrt{1+4l+4l^2+\frac{12EG^2M^2}{\hbar^2}}-1\right)} = 0
\end{equation}

This in turn implies that for the wave function to vanish:

\begin{equation}
\left(\sqrt{1+4l+4l^2+\frac{12EG^2M^2}{\hbar^2}}\right)  > 1
\end{equation}

This is the wave function vanishing condition in terms of the quantum number $l$.

Since $R(0) = 0$ when the above condition is satisfied, by Eq. (14):
\begin{equation}
\psi(r,\theta,\phi) = 0
\end{equation}

\section{Conclusion}
\label{conclusion}

We have explicitly shown through the Hamiltonian obtained from cometric of The Schwarzschild spacetime, a quantum wave function vanishes if: 
\begin{equation}
\left(\sqrt{1+4l+4l^2+\frac{12EG^2M^2}{\hbar^2}}\right)  > 1
\end{equation}

This has potentially deep implications regarding the conservation of quantum information with respect to the evolution of a \emph{universal} wave function.

\end{document}